\documentclass[a4paper]{appolb}

\usepackage{orcidlink}

\usepackage{cite}
\usepackage[initials,citation-order,nobysame]{amsrefs}

\BibSpec{article}{
+{} {\PrintAuthors} {author}
+{.} { } {part}
+{:} { \textit} {subtitle}
+{,} { \PrintContributions} {contribution}
+{.} { \PrintPartials} {partial}
+{,} { } {journal}
+{} { \textbf} {volume}
+{} { \PrintDatePV} {date}
+{,} { \issuetext} {number}
+{,} { \eprintpages} {pages}
+{,} { } {status}
+{,} { \PrintDOI} {doi}
+{,} { \eprint} {eprint}
+{} { \parenthesize} {language}
+{} { \PrintTranslation} {translation}
+{;} { \PrintReprint} {reprint}
+{.} { } {note}
+{.} {} {transition}
+{} {\SentenceSpace \PrintReviews} {review}
}

\usepackage[font=scriptsize]{caption}

\usepackage{verbatim}

\usepackage{float}
\usepackage{subfloat}

\usepackage{mathtools}

\usepackage{subcaption}

\usepackage{ragged2e}
\usepackage{amsmath}
\usepackage{amsfonts}
\usepackage{amssymb}
\usepackage{graphicx}
\usepackage{slashed}
\usepackage{bm}
\usepackage{color}
\usepackage{epsf}
 \usepackage{orcidlink}
\usepackage{psfrag}
\usepackage{appendix}

\usepackage[parfill]{parskip}
\usepackage{hyperref}
\hypersetup{
  colorlinks   = true, 
  urlcolor     = black, 
  linkcolor    = black, 
  citecolor   = red 
}

\usepackage{vmargin}
\setpapersize{A4}
\setmargins{2.5cm}       
{1.5cm}                        
{16.5cm}                      
{23.42cm}                    
{10pt}                           
{1cm}                           
{0pt}                             
{2cm}                           

\newcommand{\bq}{\bar{q}} 
\newcommand{\bp}{\bar{p}} 
\newcommand{\thetaL}{\theta_\ell} 
\newcommand{\phiL}{\phi_\ell} 
 
\newcommand{\cffh}{\mathcal{H}} 

\begin{document}
\title{\vspace{-3cm}{\bf Prospects for GPDs extraction with Double DVCS}%
\thanks{Presented by V.M.F. at the XXIX Cracow Epiphany Conference (Poland), on January 16th, 2023.}%
}
\author{K.~Deja\,\orcidlink{0000-0002-9083-2382}, V.~Mart\'inez-Fern\'andez\,\orcidlink{0000-0002-0581-7154}, P.~Sznajder\,\orcidlink{0000-0002-2684-803X}, J.~Wagner\,\orcidlink{0000-0001-8335-7096}
\address{National Centre for Nuclear Research (NCBJ), 02-093 Warsaw, Poland}
\\[3mm]
{B.~Pire\,\orcidlink{0000-0003-4882-7800}
\address{Centre de Physique Th\'eorique, CNRS, École Polytechnique, I.P. Paris, 91128 Palaiseau, France}
}
}
\maketitle

\begin{abstract}
Double deeply virtual Compton scattering (DDVCS) is the process where an electron scatters off a nucleon and produces a lepton pair. The main advantage of this process in contrast with deeply virtual and timelike Compton scatterings (DVCS and TCS) is the possibility of directly measuring GPDs for $x\neq\pm\xi$ at leading order in $\alpha_s$ (LO). We present a new calculation of the DDVCS amplitude based on the methods developed by R.~Kleiss and W.~J.~Stirling in the 1980s. These techniques produce expressions for amplitudes that are perfectly suited for implementation in numerical simulations. Via the PARTONS software, the correctness of this new formulation has been tested by comparing the DVCS and TCS limits of DDVCS with independent calculations of DVCS and TCS.
\end{abstract}

\section{Introduction}
Generalized parton distributions (GPDs) \cite{Diehl:2003ny,Belitsky:2005qn} are off-forward matrix elements of quark and gluon operators that represent a 3D version of the usual parton distribution functions (PDFs). While PDFs are accessible in inclusive processes (out of which deep inelastic scattering off the nucleon, DIS, is the golden channel), GPDs appear in exclusive processes such as deeply virtual and timelike Compton scattering (DVCS - the golden channel for GPDs - and TCS), and double deeply virtual Compton scattering (DDVCS).

\begin{figure}[htb]
\centerline{
\includegraphics[scale=0.8]{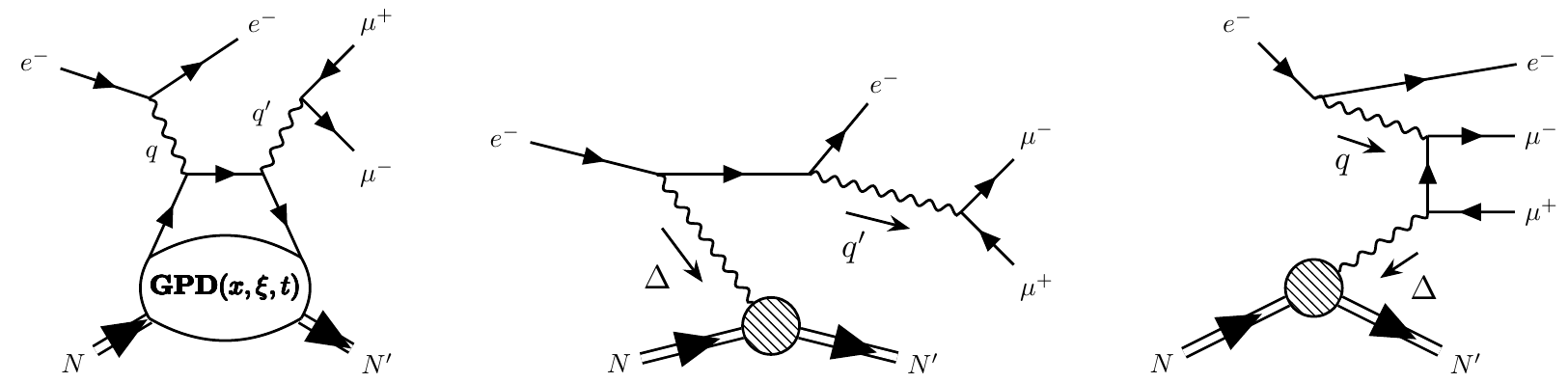}}
\caption{DDVCS (left) and BH diagrams which are denoted as BH1 (middle) and BH2 (right). Crossed-counterparts are not included.}
\label{fig:ddvcs_and_bh}
\end{figure}

To access GPDs in DDVCS \cite{Belitsky:2002tf, guidal2003} one has to consider the exclusive electroproduction of a lepton pair,
\begin{equation}
    e(k) + N(p) \to e'(k') + N'(p') + \mu^+(\ell_+) + \mu^-(\ell_-) \,,
    \label{reaction}
\end{equation}
which receives contributions not only from pure DDVCS, but also from a QED background known as Bethe-Heitler (BH) sub-process, vid.~Fig.~\ref{fig:ddvcs_and_bh}.

Present interest in DDVCS is rooted on the possibility of directly accessing GPDs in the region $x\neq\pm\xi$ in a leading order (LO) analysis. This is a consequence of the existence of two virtualities $Q^2 = -(k-k')^2$ and $Q^{\prime 2} = (\ell_+ + \ell_-)^2$ which modifies the coefficient function that convolutes with the GPD with respect to DVCS and TCS. In terms of the skewness $\xi$ and the {\it generalized} Bj{\"o}rken variable $\rho$,
\begin{equation}
    \xi = \frac{-\Delta\bq}{2\bp\bq},\quad \rho = \frac{-\bq^2}{2\bp\bq}\,,
\end{equation}
where $\bp = (p+p')/2$, $\bq = (q+q')/2$ and $\Delta = p'-p$, the DDVCS amplitude depends on the GPDs via the Compton form factors (CFFs):
\begin{equation}\label{A_ddvcs}
    {\rm CFF} \sim {\rm PV}\left( \int_{-1}^1 dx\frac{1}{x - \rho} {\rm GPD}(x, \xi, t)\right) - \int_{-1}^1 dx\ i\pi\delta(x - \rho){\rm GPD}(x, \xi, t) \pm \cdots\,.
\end{equation}
Here, the $+$ ($-$) sign corresponds to axial (vector) GPDs,  the ellipses accounts for $x\rightarrow -x$ terms, ${\rm PV}$ stands for Cauchy's principal value and $t = \Delta^2$ is the usual Mandelstam variable. As a result, one can measure GPDs for $x = \rho$ for which $\rho \neq \pm\xi$ as long as both vitualities $Q^2, Q^{\prime 2}$ are non-zero. This is different to DVCS case\footnote{Or TCS with $\xi\rightarrow -\xi$} for which the CFFs in the amplitude enter as in Eq.~(\ref{A_ddvcs}) with $\rho\rightarrow\xi$. This restricts the LO study of GPDs to the line $x = \xi$.

Although a quite detailed study of the phenomenological peculiarities of DDVCS already exists \cite{Belitsky:2003fj}, we revisit this process in the view of the near future experiments at both fixed target facilities \cite{Chen:2014psa,Camsonne:2017yux,Zhao:2021zsm} and electron-ion colliders \cite{AbdulKhalek:2021gbh, Anderle:2021wcy}. For this purpose we \cite{Deja:2023ahc} make use of Kleiss-Stirling (KS) techniques \cite{Kleiss:1984dp, Kleiss:1985yh}, which deals directly with the amplitude and render expressions that are perfectly suited for implementation in PARTONS platform \cite{Berthou:2015oaw} and so for phenomenological studies.

\section{Formulation {\`a} la Kleiss-Stirling}

In 1980s, Kleiss and Stirling developed some spinor techniques to compute  scattering amplitudes as an alternative to the usual approach based on dealing with traces of Dirac-gamma matrices. In that regard, the following products of spinors for two light-like vectors $a$ and $b$ become the building blocks of the amplitudes and define two scalars ($\pm$ stand for helicities):
\begin{align}
    s(a, b) & = \bar{u}(a,+)u(b,-) = -s(b, a)\,, \label{sKS_def}\\
    t(a, b) & = \bar{u}(a,-)u(b,+) = [s(b, a)]^*\,. \label{tKS_def}
\end{align}

Explicit computation of these bilinears show that $s(a, b)$ acquires the simple form:
\begin{equation}\label{sKS_expression}
    s(a, b) = (a^2 + ia^3)\sqrt{\frac{b^0 - b^1}{a^0 - a^1}} - (a\leftrightarrow b)\,,
\end{equation}
as long as $a\cdot\kappa_0\neq 0$ and $b\cdot\kappa_0 \neq 0$ with $\kappa^\mu_0 = (1, 1, 0, 0)$.

In turn, we can define two functions that will play a key role on the computation: the contraction of two currents
\begin{align}\label{function_f}
    f(\lambda, k_0, k_1; \lambda', k_2, k_3) = & \bar{u}(k_0,\lambda)\gamma^\mu u(k_1, \lambda)\bar{u}(k_2,\lambda')\gamma_\mu u(k_3,\lambda') \nonumber\\
    = & 2 [ s(k_2,k_1)t(k_0,k_3)\delta_{\lambda-}\delta_{\lambda'+} + t(k_2,k_1)s(k_0,k_3)\delta_{\lambda+}\delta_{\lambda'-} \nonumber\\
    & + s(k_2,k_0)t(k_1,k_3)\delta_{\lambda+}\delta_{\lambda'+} + t(k_2,k_0)s(k_1,k_3)\delta_{\lambda-}\delta_{\lambda'-} ]\,,
\end{align}
and the contraction of a current with a light-like vector $a$:
\begin{equation}\label{function_g}
    g(s, \ell, a, k) = \bar{u}(\ell, s)\slashed{a}u(k, s) 
    = \delta_{s+}s(\ell,a)t(a,k) + \delta_{s-}t(\ell,a)s(a,k)\,.
\end{equation}

\subsection{Example: BH1 {\`a} la KS}
Making use of the quantities defined above, for the case of the middle diagram in Fig.~\ref{fig:ddvcs_and_bh} which corresponds to the first BH contribution, namely BH1, the amplitude of this sub-process reads (up to propagators and factor $ie^4$):
\begin{align}\label{iwideM_BH1}
    i\mathcal{\widetilde{M}}_{{\rm BH1}} = & (F_1+F_2)\sum_L f(s_\ell,\ell_-,\ell_+; s, k', L)\Big( Y_{s_2s_1}f(s,L,k; +,r'_{s_2}, r_{s_1}) + Z_{s_2s_1}f(s,L,k; -,r'_{-s_2}, r_{-s_1}) \Big) \nonumber\\ 
    & - \frac{F_2}{2M}J^{(2)}_{s_2 s_1}\sum_{L, R}f(s_\ell, \ell_-, \ell_+; s, k', L)g(s, L, R, k)\,
\end{align}
where $F_1, F_2$ are the electromagnetic form factors, $M$ is the target mass and $J^{(2)}$ is a combination of scalars in Eqs.~(\ref{sKS_def}) and (\ref{tKS_def}) dependent on the spin and momentum of the target in its final ($s_2, p'$) and initial ($s_1, p$) states. $Y, Z$ are complex phases dependent on the target states too. Sums run over two sets of light-like momenta\footnote{Electron, muon and antimuon are considered massless.}, namely $L\in\{k',\ell_-,\ell_+\}$ and $R\in\{r_1, r_2, r'_1, r'_2\}$ where $p = r_1+r_2$ and $p'=r'_1+r'_2$.
Eq.~(\ref{iwideM_BH1}) can be clearly interpreted in terms of contractions of leptonic and hadronic currents as well as momenta via definitions (\ref{function_f}) and (\ref{function_g}).

\section{DVCS and TCS limits}
In this section we numerically check our results against DVCS and TCS limits, which were previously worked out in \cite{BELITSKY2014214,Berger:2001xd} and implemented in the PARTONS framework. In these tests we consider Goloskokov-Kroll GPD model, see for example Ref.~\cite{Goloskokov_2007}, the renormalization and factorization scales are $\mu_R^2 = \mu_F^2 = Q^2+Q'^2$, while the skewness and generalized Bjorken variables are evaluated at $t = 0$, which is equivalent to drop terms proportional to $t/(Q^2+Q^{\prime 2})$. Without denying the importance of NLO corrections to the amplitudes \cite{Pire:2011st,Moutarde:2013qs}, we stay at the Born order level for the time being. The cross-sections are given as a function of seven variables out of which three are angles: $\phi$ which describes the azimuthal direction of the final-state hadron with respect to the electron beam plane, and $\phiL$ and $\thetaL$ which represent azimuthal and polar orientations of the muon in the produced lepton pair center of mass frame, respectively. The first one is given according to Trento's convention \cite{trento}, while the other two are considered in BDP frame \cite{Berger:2001xd}.

\begin{figure}[htb]
\centerline{
\includegraphics[scale=0.45]{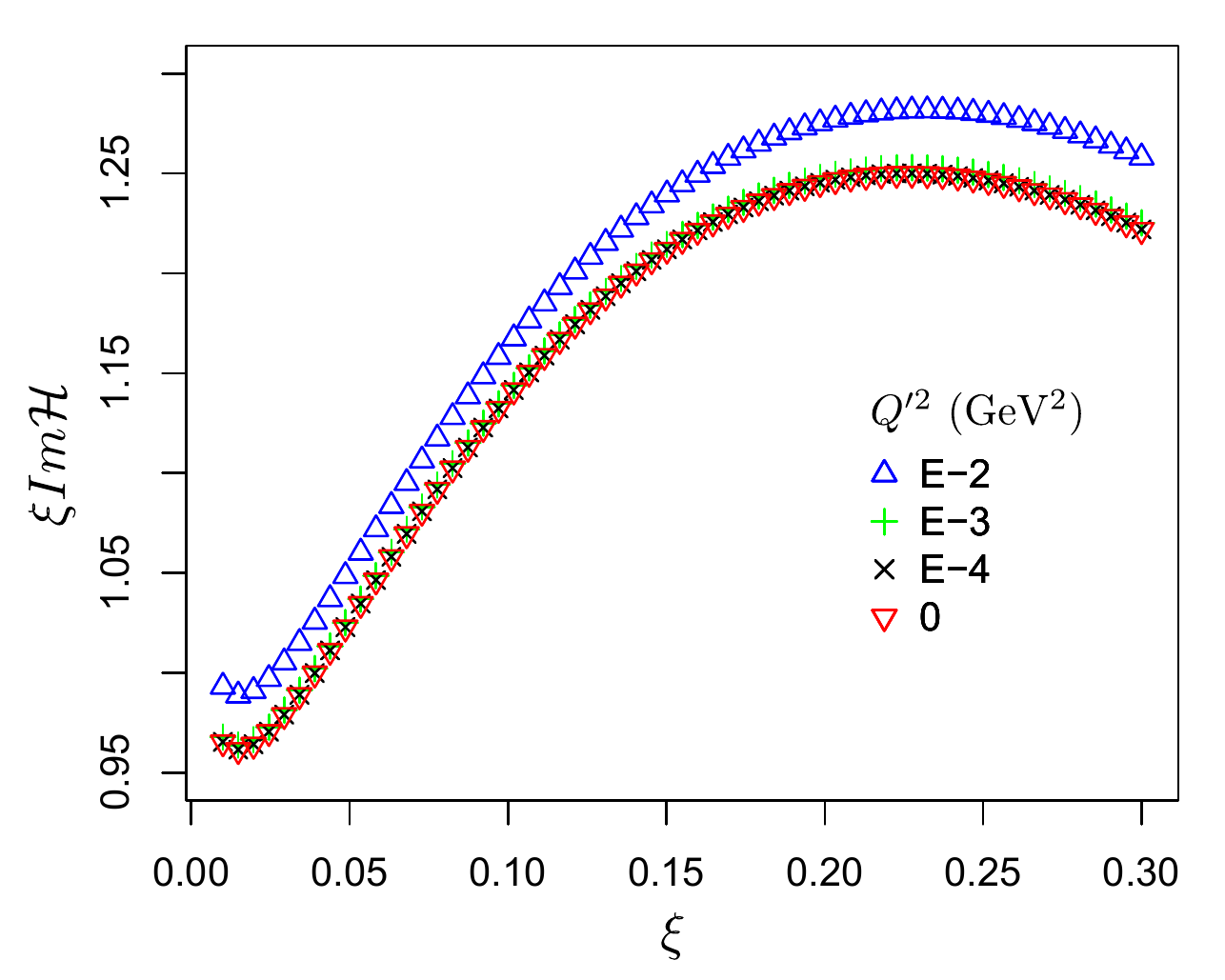}
\includegraphics[scale=0.45]{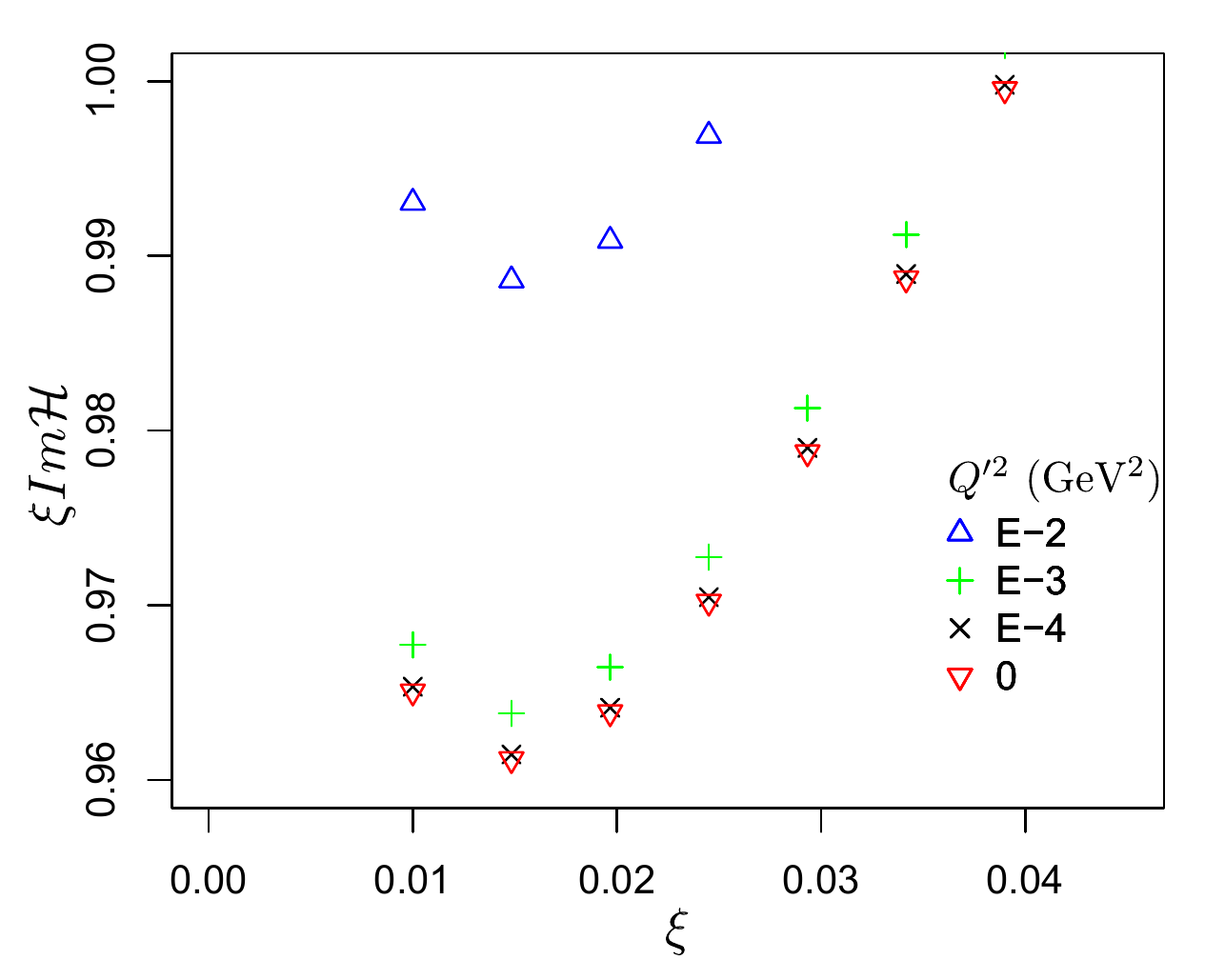}}
\caption{Left: Imaginary part of CFF $\cffh(\xi, t = -0.15\mathrm{\ GeV^2})$ as a function of the skewness $\xi$ for subsequently smaller values of $Q^{\prime 2}$ and comparison with DVCS  CFF $\cffh$ ($Q^{\prime 2} = 0$). Value of the spacelike virtuality is taken to be $Q^2 = 1.5$ GeV$^2$. Right: zoom in the region $\xi\in(0.01, 0.04)$.}
\label{fig:xiImH_vs_xi}
\end{figure}

In Fig.~\ref{fig:xiImH_vs_xi}, CFF $\cffh$ is depicted as a function of $\xi$ as it reaches the DVCS limit ($Q^{\prime 2} = 0$). Points corresponding to proper DVCS are computed with an independent code available in PARTONS and, as it is shown, DDVCS' CFF $\cffh$ approaches DVCS value without discontinuities. Same conclusion is reached when TCS limit and other CFFs are considered. 

As $Q^{\prime 2}\rightarrow 0$, DVCS cross-section is recovered from DDVCS. The relation between these two processes comes as
\begin{equation}
    \int d\Omega_\ell \underbrace{\frac{d^7\sigma}{dx_B dQ^2 dQ^{\prime 2}d|t|d\phi d\Omega_\ell}}_{{\rm DDVCS}} \xrightarrow{Q^{\prime 2} \rightarrow 0} \underbrace{\left( \frac{d^4\sigma}{dx_B dQ^2 d|t|d\phi} \right)}_{{\rm DVCS}} \frac{\mathcal{N}}{Q^{\prime 2}} \,,
    \label{eq:limitPrescriptionDVCS}
\end{equation}
where we have integrated-out the lepton pair and accounted for the splitting of the outgoing virtual photon into the pair via the factor $\mathcal{N} = \alpha_{\rm em}/(3\pi)$ \cite{Vanderhaeghen_1999}. Prescription (\ref{eq:limitPrescriptionDVCS}) holds also for BH1 contribution.

The same way, as $Q^2\rightarrow 0$ DDVCS tends to TCS. In this case, one needs to consider the photon flux, $\Gamma$, calculated under the equivalent-photon approximation (EPA) \cite{kessler_epa, kessler_halArchives} and integrate over $\phi$:
\begin{equation}
    \int d\phi \underbrace{\frac{d^7\sigma}{dx_B dQ^2 dQ^{\prime 2}d|t|d\phi d\Omega_\ell}}_{{\rm DDVCS}} \xrightarrow{Q^{2} \rightarrow 0} \underbrace{\left( \frac{d^4\sigma}{dQ^{\prime 2}d|t|d\Omega_\ell} \right)}_{{\rm TCS}} \frac{d^2\Gamma}{dx_{B}dQ^2} \,,
    \label{eq:limitPrescriptionTCS}
\end{equation}
where
\begin{equation}
    \frac{d^2\Gamma}{dx_{B}dQ^2} = \frac{\alpha_{\rm em}}{2\pi Q^2}\left( 1 + \frac{(1-y)^2}{y} - \frac{2(1-y)Q_{\rm min}^2}{yQ^2} \right)\frac{\nu}{Ex_B} \,.
\end{equation}
Here,
\begin{equation}
    \nu = \frac{Q^2}{2 M x_B}\quad \mathrm{and}\quad Q^2_{\rm min} = \frac{(ym_e)^2}{1-y}
\end{equation}
are the energy carried away by the incoming virtual photon and the minimum value of the spacelike virtuality for which $m_e$ is the electron mass, respectively. Prescription (\ref{eq:limitPrescriptionTCS}) holds also for BH2 contribution.

\begin{figure}[htb]
\centering
\includegraphics[scale=0.45]{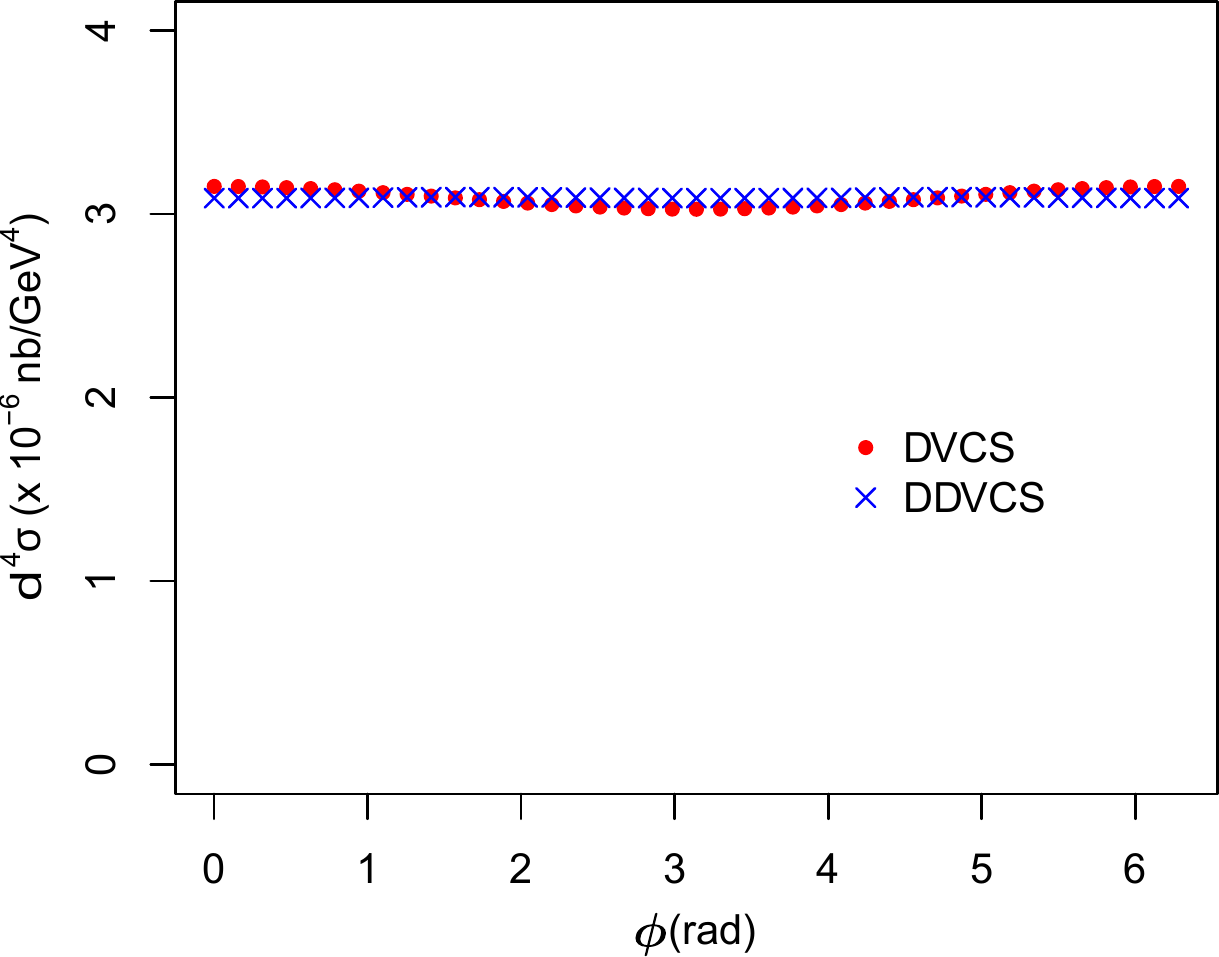}
\includegraphics[scale=0.45]{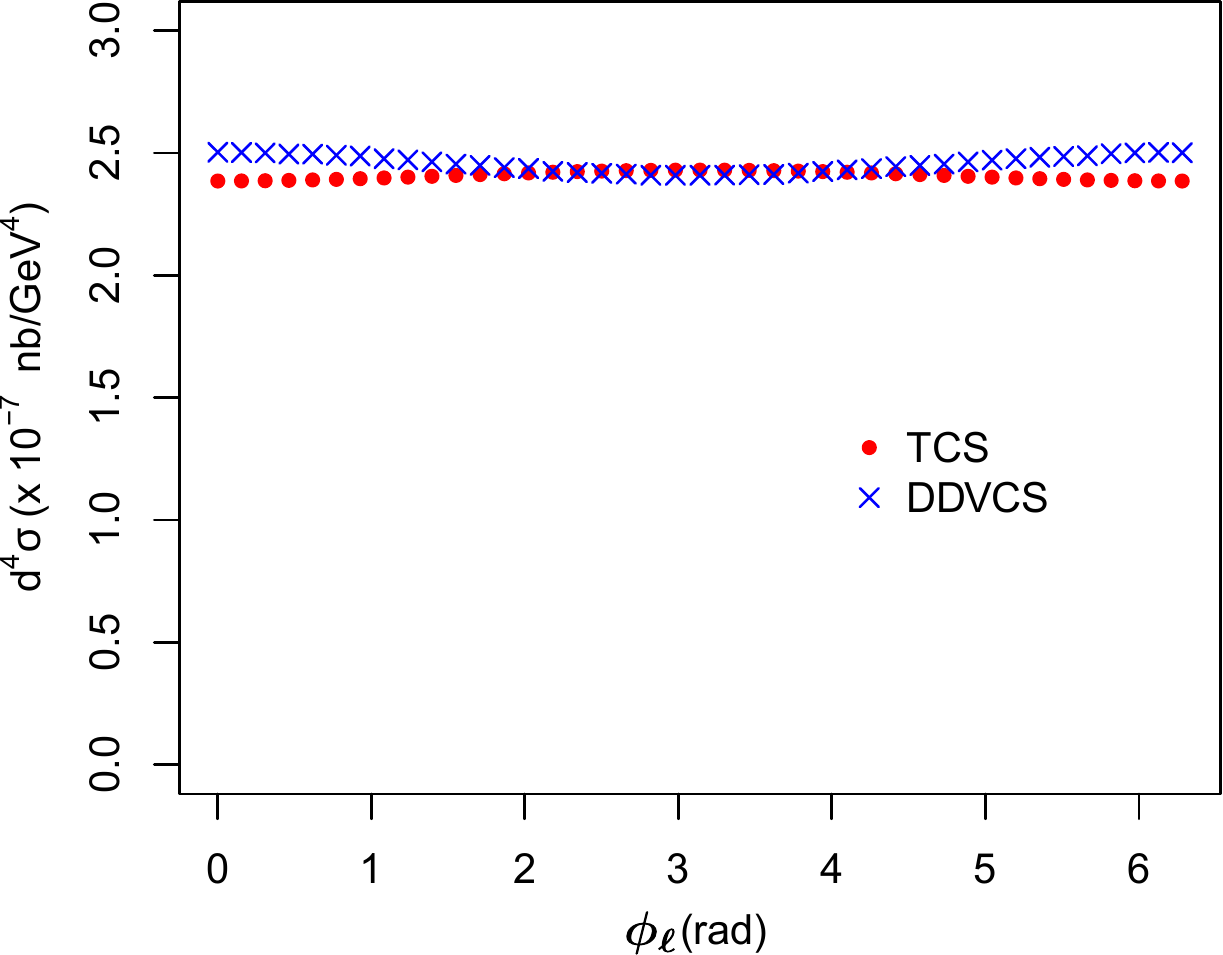}
\caption{Pure DVCS (left) and TCS (right) contributions to the cross-section  versus the corresponding limit of DDVCS. For left plot: $x_B = 0.2$, $t = -0.25$ GeV$^2$, $Q^2 = 40$ GeV$^2$ and incoming electron beam energy $E = 160$ GeV. For right plot: $x_B = 10^{-4}$, $t = -0.25$ GeV$^2$, $Q^{\prime 2} = 33$ GeV$^2$, $\thetaL = 1.04\pi/4$ rad and $E = 160$ GeV.}
\label{fig:dvcsAndTcsLim}
\end{figure}

\begin{figure}[htb]
\centering
\includegraphics[scale=0.47]{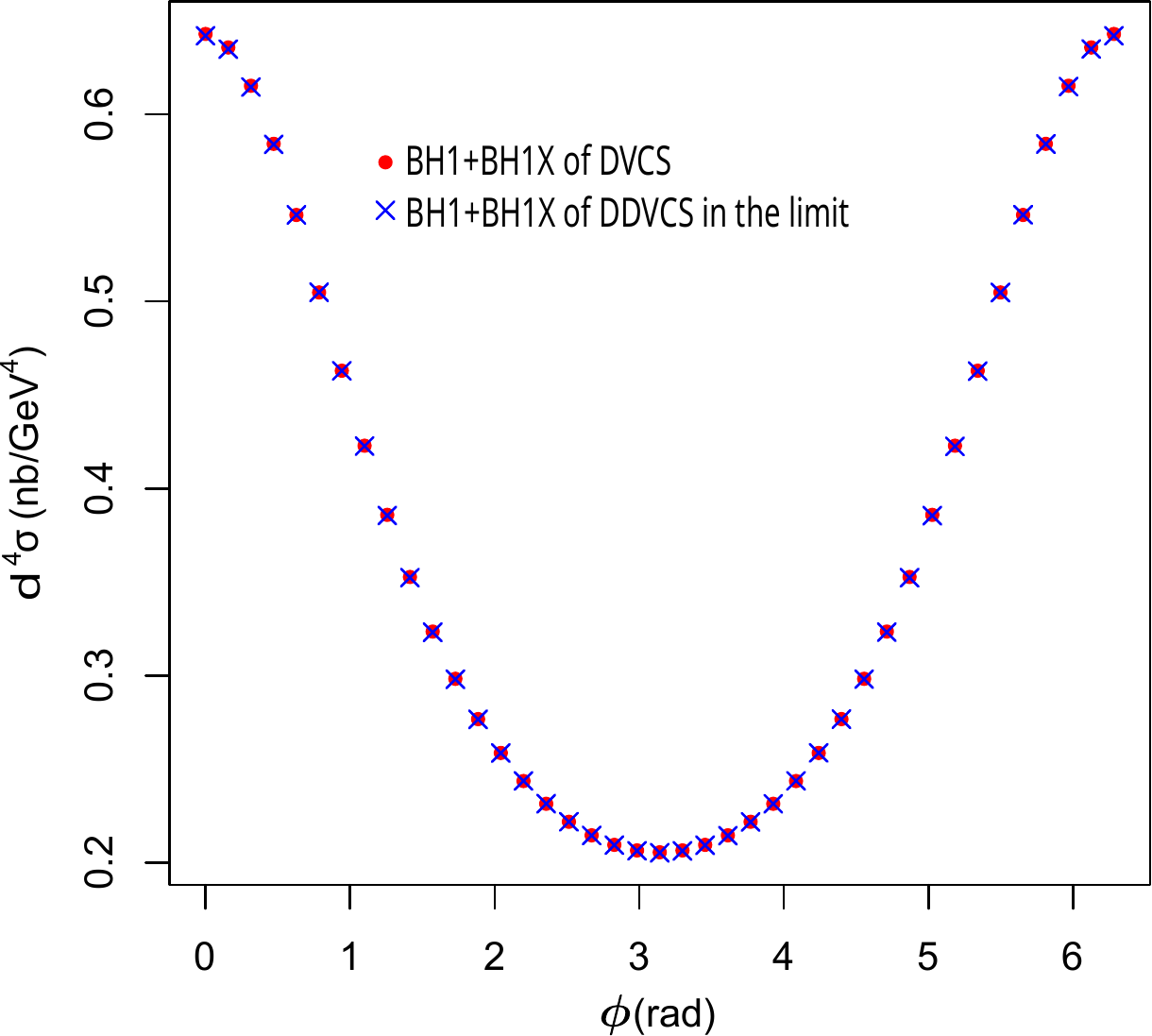}
\includegraphics[scale=0.45]{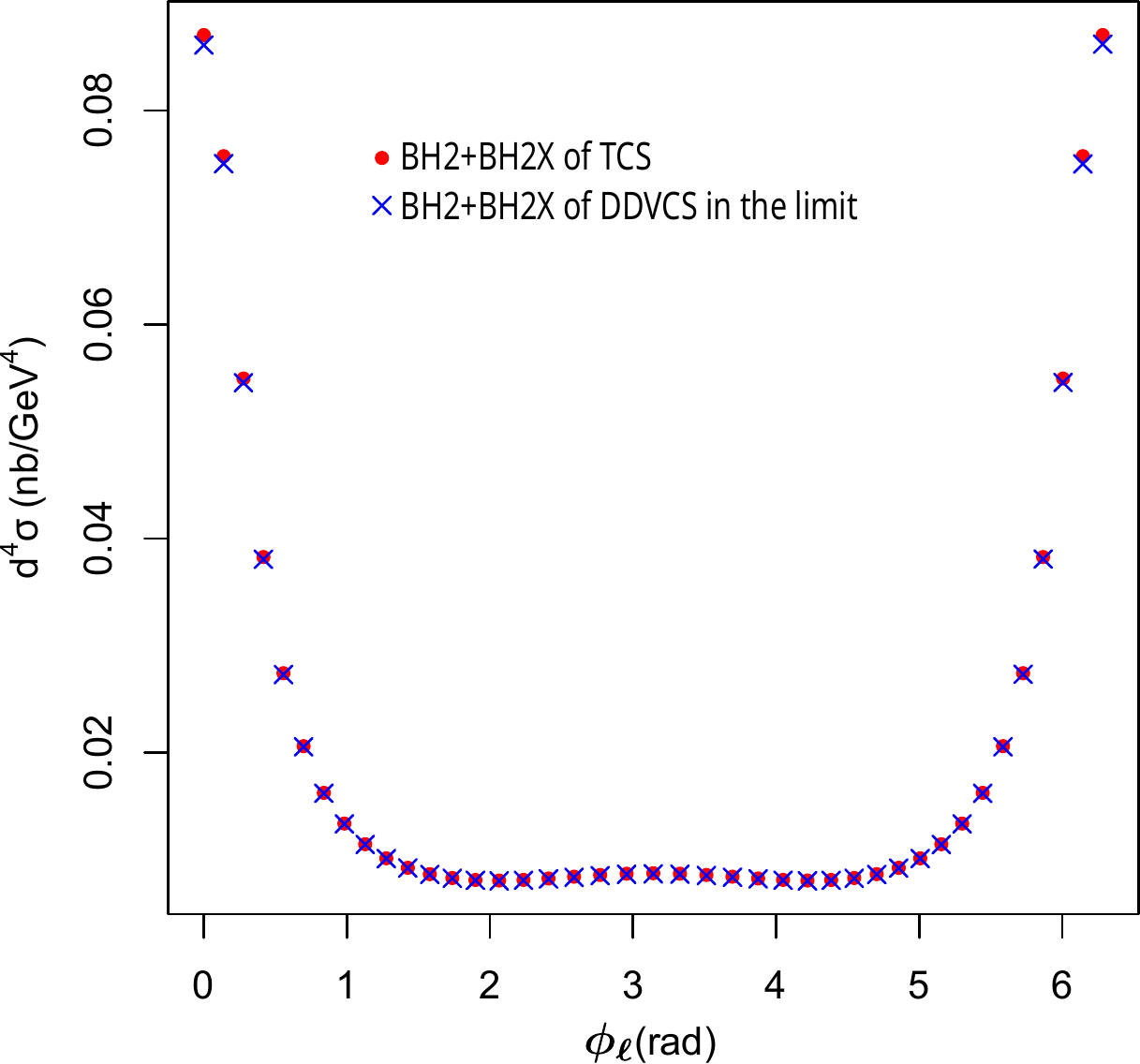}
\caption{BH contribution to the DVCS (left) and TCS (right) cross-sections versus the corresponding limit of DDVCS. For left plot: $x_B = 0.04$, $t = -0.1$ GeV$^2$, $Q^2 = 10$ GeV$^2$ and incoming electron beam energy $E = 160$ GeV. For right plot: $x_B = 10^{-4}$, $t = -0.1$ GeV$^2$, $Q^{\prime 2} = 3$ GeV$^2$, $\thetaL = 1.04\pi/4$ rad and $E = 160$ GeV.}
\label{fig:bhInTheLim}
\end{figure}

The comparison for cross-sections is shown in Fig.~\ref{fig:dvcsAndTcsLim} for DDVCS against DVCS and TCS, and in Fig.~\ref{fig:bhInTheLim} for the BH backgrounds. Also here, proper DVCS and TCS processes are evaluated with independent codes available in PARTONS. These codes are numerical implementations of works published in Refs.~\cite{BELITSKY2014214} and \cite{Berger:2001xd}.

Testing for different ratios of $|t|/Q^2$ and $|t|/Q^{\prime 2}$, we conclude that the slight disagreement observed in Fig.~\ref{fig:dvcsAndTcsLim} is the result of kinematical higher-twists. These effects come from the different frames used to describe DDVCS, DVCS and TCS. Because there is no twist expansion for pure QED processes, BH contributions in Fig.~\ref{fig:bhInTheLim} show a perfect matching.

As a final remark, on top of these consistency checks, we are working on the predictions for cross-sections and asymmetries assessing the measurability of DDVCS in both current (JLab12) and future experiments (JLab20+, EIC) \cite{Deja:2023ahc}.

\scriptsize{{\bf Acknowledgements.} The works of J.W. are  supported by the grant 2017/26/M/ST2/01074 of the National Science Center (NCN), Poland. Development of EpIC Monte Carlo generator by P.S. was supported by the grant 2019/35/D/ST2/00272 of the NCN. This work is also partly supported by the COPIN-IN2P3 and by the European Union’s Horizon 2020 research and innovation programme under grant agreement No 824093. The works of V.M.F.~are supported by PRELUDIUM grant 2021/41/N/ST2/00310 of the NCN.}

\bibliographystyle{amsrefs}
\bibliography{main.bib}
\end{document}